\documentstyle[11pt]{article}

\evensidemargin=\oddsidemargin
\def\lumdens{\hbox{$\cal L$}}

\def\beginfigures{\newpage \section*{Figures} \begin{enumerate}}
\def\endfigures{\end{enumerate}}
\def\figure#1{\item \label{#1}}

\def\etal{{\em et~al.}}
\def\sci#1{\ifmmode \times 10^{#1} \else $\times 10^{#1}$\fi}

\def\unit#1{\ifmmode \;{\rm #1}\else $\;{\rm #1}$\fi}
\def\Mpc{\unit{Mpc}}

\def\Gyr{\unit{Gyr}}
\def\Myr{\unit{Myr}}

\def\mcommand#1{\ifmmode #1\else $#1$\fi}
\def\OmegaB{\mcommand{\Omega_{\rm baryon}}}
\def\OMatter{\mcommand{\Omega_{\rm matter}}}
\def\rhoB{\mcommand{\rho_{\rm baryon}}}
\def\MB{\mcommand{M_{\rm baryon}}}
\def\OmegaBC{\mcommand{\Omega_{\rm B,C}}}
\def\OmegaVis{\mcommand{\Omega_{\rm vis}}}
\def\rhoVis{\mcommand{\rho_{\rm vis}}}
\def\OmegaClus{\mcommand{\Omega_{\rm cluster}}}
\def\Hunits{\mcommand{\;{\rm km/sec/Mpc}}}
\def\rhocrit{\mcommand{\rho_{\rm crit}}}

\def\HOmega{\mcommand{H_0\hbox{-}\Omega}}

\def\elem#1#2{\mcommand{{}^{#1}{\rm #2}}}
\def\element#1#2{\mcommand{{}^{#2}{\rm #1}}}
\def\He#1{\element{He}{#1}}

\def\be{\begin{equation}}
\def\ee{\end{equation}}

\def\gtsim{\mathrel{\mathpalette\fun >}}
\def\fun#1#2{\lower3.6pt\vbox{\baselineskip0pt\lineskip.9pt
  \ialign{$\mathsurround=0pt#1\hfil##\hfil$\crcr#2\crcr\sim\crcr}}}

\def\thefootnote{\if\value{footnote} > 9 \setcounter{footnote}{1} \fi
    \hbox{$\fnsymbol{footnote}$}}

\makeatletter

\let\old@bibitem=\@bibitem
\def\@bibitem#1#2{\@ifundefined{r@#1}{}{\@warning
  {Multiple entries for reference `#1'}}\@ifundefined{b@#1}{\@warning
  {Reference `#1' not cited}}{\global\@namedef{r@#1}{#2}}}
\let\old@lbibitem=\@lbibitem
\def\@lbibitem[#1]#2#3{\@ifundefined{r@#2}{}{\@warning
  {Multiple entries for reference `#2'}}\@ifundefined{b@#2}{\@warning
  {Reference `#2' not cited}}{\global\@namedef{r@#2}{#3}
  \global\@namedef{s@#2}{#1}}}

\def\@putbibrefs#1{\expandafter\@makearef #1*,}
\def\@makearef#1,{\if*#1 \let\next=\relax \else \let\next=\@makearef
  \@ifundefined{s@#1}{\bibitem{#1}}{\bibitem[\@nameuse{s@#1}]{#1}}
  \@ifundefined{r@#1}{Citation label `#1'?}{\@nameuse{r@#1}}
  \fi \next}

\def\endthebibliography{\let\@bibitem=\old@bibitem \let\@lbibitem=\old@lbibitem
  \@putbibrefs\@bibrefslist \endlist}

\def\@citex[#1]#2{\if@filesw\immediate\write\@auxout{\string\citation{#2}}\fi
  \def\@citea{}\@cite{\@for\@citeb:=#2\do
    {\@citea\def\@citea{,\penalty\@m}\@ifundefined
       {b@\@citeb}{{\bf ?}\@warning
       {Citation `\@citeb' on page \thepage \space undefined}}%
\hbox{\csname b@\@citeb\endcsname}\@ifundefined
       {d@\@citeb}{\global\@namedef{d@\@citeb}{y}\@addlabel{\@citeb}}{}}}{#1}}

\def\nocite#1{\@ifundefined{d@#1}{\@addlabel{#1}}{\@warning
	{Label `#1' cited.  \string\nocite\space not necessary.}}}

\def\@addlabel#1{\@ifundefined{@bibrefslist}{\xdef\@bibrefslist{#1,}}{\xdef
	\@bibrefslist{\@bibrefslist#1,}}}
\let\old@cite=\@cite
\def\norma@cite#1#2{{\footnotesize ${}^{#1\if@tempswa,#2\fi}$}}
\def\normb@cite#1#2{{#1\if@tempswa,#2\fi}}
\def\citea#1{\global\let\@cite=\norma@cite\cite{#1}\@msf\gdef\@msf
  {\spacefactor=1000}}
\def\citeb{\global\let\@cite=\normb@cite\cite}
\def\thebibliography#1{\section*{References\@mkboth
 {REFERENCES}{REFERENCES}}\list
 {\arabic{enumi}.}{\settowidth\labelwidth{#1}\leftmargin\labelwidth
 \advance\leftmargin\labelsep
 \usecounter{enumi}}
 \def\newblock{\hskip .11em plus .33em minus .07em}
 \sloppy\clubpenalty4000\widowpenalty4000
 \sfcode`\.=1000\relax}
\def\bckup#1{\setbox0=\hbox{#1}\hbox to
  0.5\wd0{#1\xdef\@msf{\spacefactor=\the\spacefactor}\hss}}
\def\@msf{\spacefactor=1000}
\makeatother

\begin{document}

\null
\begin{center}
\bigskip
\rightline{FERMILAB--PUB--95/???-A}
\rightline{astro-ph/9504026}
\rightline{submitted to {\it Comments on Nuclear and Particle Physics}}

\vskip 0.35in
{\Large The Nuclear Impact on Cosmology: The \HOmega\ Diagram}

\vskip 0.2in
Craig J. Copi$^{1,2}$ and David N. Schramm$^{1,2,3}$

\vskip 0.2in
{\it $^1$Department of Physics\\
The University of Chicago, Chicago, IL~~60637-1433}\\

\vskip 0.1in
{\it $^2$NASA/Fermilab Astrophysics Center\\
Fermi National Accelerator Laboratory, Batavia, IL~~60510-0500}\\

\vskip 0.1in
{\it $^3$Department of Astronomy \& Astrophysics\\
Enrico Fermi Institute, The University of Chicago,
Chicago, IL~~60637-1433}\\

\medskip
\end{center}

\begin{abstract}

	The \HOmega\ diagram is resurrected to dramatically illustrate the
nature of the key problems in physical cosmology today and the role that
nuclear physics plays in many of them.  In particular it is noted that the
constraints on \OmegaB\ from big bang nucleosynthesis do not overlap with the
constraints on \OmegaVis\ nor have significant overlap with the lower bound on
$\Omega$ from cluster studies.  The former implies that the bulk of the
baryons are dark and the later is the principle argument for non-baryonic dark
matter.  A comparison with hot x-ray emitting gas in clusters is also made.
The lower bound on the age of the universe from globular cluster ages
(hydrogen burning in low mass stars) and from nucleocosmochronology also
illustrates the Hubble constant requirement $H_0 \le 66 \Hunits$ for $\Omega_0
= 1$\@.  It is also noted that high values of $H_0$ ($\sim 80\Hunits$) even
more strongly require the presence of non-baryonic dark matter.  The lower
limit on $H_0$ ($\ge 38\Hunits$) from carbon detonation driven type~Ia
supernova constrains long ages and only marginally allows \OmegaB\ to overlap
with \OmegaClus.  Diagrams of \HOmega\ for $\Lambda_0=0$ and $\Lambda_0\neq 0$
are presented to show that the need for non-baryonic dark matter is
independent of $\Lambda$.

\end{abstract}

\section{Introduction}

	In 1974, Gott, Gunn, Schramm, and Tinsley\citea{ggst}~(hereafter, GGST)
showed that a plot of the Hubble constant, $H_0$, verse the the dimensionless
density parameter,
\be \Omega \equiv \frac{\rho}{\rhocrit}, \label{eqn:Omega} \ee
where $\rho$ is the mass density and
\be \rhocrit = \frac{3 H_0^2}{8\pi G} \ee
is the critical cosmological density, well illustrated the issues in physical
cosmology, particularly for models with cosmological constant $\Lambda=0$.
Twenty years later we again use the \HOmega\ diagram and show that the
constraints of GGST have not changed significantly but the interpretation now
illustrates the critical issues in physical cosmology today, namely the dark
matter and age problems.  As nuclear/particle astrophysicists we note with
pride (or fear) how many of the most significant lines on the \HOmega\ diagram
have their origin in nuclear physics arguments.

	It will be shown how the \HOmega\ diagram dramatically illustrates
that there are at least two dark matter problems, namely the bulk of the
baryons are dark and the bulk of the matter in the universe is non-baryonic.
It will also be shown that these two dark matter problems persist regardless
of the value of $H_0$\@.  These arguments center on the big bang
nucleosynthesis~(BBN) constraints on $\Omega$ in baryons\bckup,\citea{cst}
\OmegaB.  We will also review the age constraints from  globular cluster
ages\citea{shietal} and from nucleocosmochronology\bckup.\citea{ms} We will
show with the $H_0
\hbox{-} \Omega$ diagram that $\Omega_0=1$ and the cosmological constant
$\Lambda_0=0$ requires $H_0 \le 66\Hunits$\@.  Variations in age-\HOmega\
relationships for non-zero $\Lambda$ are also discussed.  A lower bound on
$H_0 \ge 38\Hunits$ from carbon-detonation powered type~Ia
supernova\citea{abw} is also plotted.  For comparison, the information on the
baryon content of hot x-ray emitting gas in clusters based on ROSAT
measurements\citea{wnef} is also discussed.  This paper will now go through
each of the constraints in turn and generate appropriate \HOmega\ diagrams.

\section{Cosmological Model}

	The Friedmann-Robertson-Walker~(FRW) cosmological model provides a
simple physical and mathematical model for describing the large scale
structure of the universe by assuming the universe is isotropic and
homogeneous.  The smoothness of the background radiation is striking
confirmation that the universe is isotropic at a level of 1 part in $10^5$
(COBE).  The assumption of homogeneity is less straight forward to confirm,
however, measurements of peculiar velocities of galaxies on very large
scales\citea{dekel} as well as radio source count studies\citea{pstk} seem to
indicate that it is valid.  Within the FRW model the distance and time scales
can be related to the Hubble constant,
\be H_0 = \frac{\dot{R} (t_0)}{R (t_0)}, \ee
where $R(t)$ is the scale factor of the universe and $t_0$ is the present age.
As we shall see the value of the Hubble constant is still quite uncertain.
Thus in practice it is useful to introduce a dimensionless factor
\be h = \frac{H_0}{100 \Hunits} \ee
to express this uncertainty.  The geometry of the universe is encoded in
$\Omega$ (and $\Lambda$); for $\Omega < 1$ ($\Omega > 1$) and $\Lambda=0$ the
universe is opened and hyperbolic (closed and spherical) and for $\Omega = 1$
it is opened and flat.  For the most part we shall assume the vacuum has no
density nor pressure, in other words we assume $\Lambda = 0$ unless explicitly
stated to the contrary.

	In the FRW, $\Lambda=0$ model the present age of the universe is
\be t_0 = \frac{f(\Omega_0)}{H_0}, \ee
where
\be f(\Omega_0) =
\def\arraystretch{2}
  \left\{ \begin{array}{ll}
	\displaystyle \frac{\Omega_0}{2 (\Omega_0 - 1)^{3/2}}
	\left[ \cos^{-1}(2/\Omega_0 - 1) - \frac{2}{\Omega_0} \sqrt{\Omega_0 -
	1} \right], & \Omega_0 > 1 \\
	\displaystyle \frac23, & \Omega_0 = 1 \\
	\displaystyle \frac{\Omega_0}{2 (1 - \Omega_0)^{3/2}}
	\left[ \frac{2}{\Omega_0} \sqrt{1 - \Omega_0}
	- \cosh^{-1}(2/\Omega_0 - 1) \right], & \Omega_0 < 1
  \end{array} \right. ,
\ee
and $\Omega_0$ is the present value of $\Omega$.

\section{The Age of the Universe}

The most stringent limits on the age of the universe come from the age of
globular clusters and nucleocosmochronology.  The age of globular clusters is
essentially determined by the rate of hydrogen burning in low mass, metal poor
stars.  When the core of the star has been completely converted to helium the
star changes its structure and no longer lies on the main sequence of a
Hertzsprung-Russel luminosity-temperature plot.  The main sequence turnoff age
is dependent on numerous aspects of the globular clusters, such as
metallicity, helium diffusion in the stars, and the initial helium
abundance\bckup.\citea{shietal} While many groups have calculated globular
cluster ages in the $14\hbox{--} 16\Gyr$ range\bckup,\citea{gcages}
Shi\citea{shi} has shown that reasonable systematic assumptions in the
calculations could lower the ages to $12\pm2$ with a firm lower bound of
$t_0\ge 10\Gyr$.  This is also consistent with an independent study by
Chaboyer\bckup.\citea{chaboyer} This lower bound can be obtained trivially by
noting that globular clusters should be burning hydrogen less rapidly than the
sun since they have lower metallicity and a slightly lower mass.  Our sun will
exhaust the hydrogen in its core in about $10\Gyr$ (based on the hydrogen
burning rate verified by the calibrated GALLEX experiment\cite{gallex}).  Thus
globular clusters must have a lower bound on their age of $t_0\ge 10\Gyr$ as
quoted above.

Nucleocosmochronology provides information about time scales over which the
elements in the solar system were formed.  This method couples knowledge of
present day abundance ratios, production ratios, and lifetimes of long lived
radioactive nuclides.  The standard methods of determining nucloechronometric
ages rely heavily on the adopted galactic evolution model.  This can lead to
large errors in the deduced galactic age.  An alternative approach is to
employ less restrictive, model-independent nucleocosmochronology which studies
the constraints that can be made without specific reference to galactic
nucleosynthesis models.  When using radioactive decay alone only a strict
lower bound is possible\bckup.\citea{sw} In particular the mean age of the
longest lived chronometer\bckup,\footnote{Although ${}^{187}{\rm Re}$ is
longer lived in its ground state, its lifetime is dependent on its thermal
environment so is not useful for a lower bound.  It does, however, constrain
the maximum mean age\bckup.\citea{meyeretal}} \elem{232}{Th}\ relative to
\elem{238}{U}, can be used to give an extremely conservative lower
bound\cite{ms,schramm} of
$\sim8\Gyr$.  Since this bound assumes all nucleosynthesis occurs in a single
event, it is obviously too extreme.  We know that \elem{235}{U},
\elem{244}{Pu}, and
\elem{129}{I} all existed in measurable abundances when the solar system
formed 4.6\Gyr\ ago and free decay from a single production event several Gyr
earlier would not be consistent (for example, \elem{129}{I}\ has a half-life of
only 17\Myr\ and \elem{244}{Pu}\ only 82\Myr).  Thus we know the production
was more spread out than a single event.  Meyer and Schramm\citea{ms}
quantified this spreading out to show that the lower bound from chronology was
$\gtsim 9\Gyr$ and subsequent analysis by Schramm\citea{schramm} using
improved limits on the production ratios pushed the bound up to $t_{\rm NC}
\ge 9.7\Gyr$.

The results from globular cluster and nucleocosmochronology provide a
consistent lower bound for the age of the universe.  We note that globular
cluster ages and nucleocosmochronology do not provide a strong upper bound to
the age of the universe since one could in principle add several Gyr of
formation time to any such age determination.  The lower bound does not have
these problems since the extreme limit is globular cluster formation on a
Kelvin-Helmholtz collapse time scale, $t\sim 10^7\unit{yr}$ at recombination,
$t \sim 10^5\unit{yr}$ which yields formation times $\ll 1\Gyr$ after the big
bang.  It is not possible to form globular clusters earlier than this time.
Based on the above results the age of the universe is constrained to be
\be t_0\ge 10 \Gyr. \ee
The resulting excluded region in the \HOmega\ plane is shown in
figure~\ref{fig:main-fig} for $\Lambda=0$, figure~\ref{fig:Lambda0.4} for
$\Omega_\Lambda=0.4$, and figure~\ref{fig:Lambda0.8} for $\Omega_\Lambda=0.8$.
Here $\Omega_\Lambda$ is defined as above~(\ref{eqn:Omega}) with
$\rho_\Lambda=\Lambda/8\pi G$.

\section{Big Bang Nucleosynthesis Limits}

	Standard homogeneous BBN accurately predicts the primordial abundances
of the light elements over 9 orders of magnitude in terms of a single
parameter, the density of baryons, $\rhoB$.  For the constraints on \OmegaB\
we adopt the recent determination by Copi, Schramm, and
Turner\bckup,\citea{cst}
\be 0.01 \le \OmegaB h^2 \le 0.02. \label{eqn:bbn-constraint} \ee
Note that $\rhoB$ is independent of the Hubble constant, thus the Hubble
constant enters into $\OmegaB$ only through \rhocrit.  The curves defined by
this choice and the excluded regions are shown in figure~\ref{fig:main-fig}.
Attempts to by-pass these constraints with inhomogeneous models have been
shown to fail in that the constraints on the light element abundances
yield essentially the same constraints on $\OmegaB$\bckup.\citea{ibbn}
Recently
Tytler and Fann\citea{tf} have observed deuterium in a quasar absorption system
that, if confirmed, restricts the baryon density to an extremely tight range
near our quoted upper limit\bckup.\citea{sc}

	It might be noted that one significant difference between our \HOmega\
diagram and that of GGST is that in 1974 BBN only provided an upper bound to
\rhoB\ from deuterium\bckup,\citea{rafs} whereas now we also have a lower bound
on \rhoB\ from \He3 plus deuterium arguments and we have the strict lithium
constraints adding consistency to the picture.

\section{Direct Measurements of $\Omega$}

\subsection{Visible Matter}

The most straight forward method of estimating $\Omega$ is to measure the
luminosity of stars in galaxies and estimate the mass to light ratio, $M/L$.
The mass density of visible objects is given by
\be \rhoVis = \frac{M}{L} \lumdens, \ee
where \lumdens\ is the average luminosity density in the wavelength band used
to determine $M/L$.  For our galaxy\citea{fg,peebles}
\be \lumdens = (1.0\pm0.3) h \sci8 L_\odot \Mpc^{-3} \quad {\rm and} \quad
\frac ML = (6 \pm 3) \frac{M_\odot}{L_\odot}. \ee
Here $M_\odot$ is the mass of the sun and $L_\odot$ is the luminosity of the
sun.  These values give
\be 0.002 \le \OmegaVis h \le 0.006. \ee
This value agrees well with the determination of $M/L$ using the visible part
of external galaxies\bckup.\citea{peebles}  The curves defined by this range
and the excluded regions are shown in figure~\ref{fig:main-fig}.

\subsection{Dynamical Measurements}

Numerous methods for dynamically measuring the density of the universe have
been developed all of which give complimentary results.  A review of many of
these methods can be found in Peebles\bckup.\citea{peebles} We shall highlight
a few of these methods.

The simplest means of measuring mass inside a radius, $r$, is via Kepler's
third law
\be GM(r) \approx v^2 r. \ee
If the mass were solely associated with the light then we would expect
$v\propto r^{-1/2}$ for some object outside of the core of the galaxy.
Instead it is observed that $v \approx {\rm constant}$ for objects far from
the center of the galaxy.  This indicates that dark matter exists in a halo
around the galaxy.  Typical estimates of the mass in halos from this method
gives $\Omega_{\rm halo} \approx 0.05$.  This dark-matter could in principle
be dark baryons; however, see the discussion on MACHOS below.  Note that
estimates of the mass density from dynamics scale with $h^2$ as does \rhocrit,
thus $\Omega$ is independent of $H_0$.

Measurements of average velocity dispersion and average separation of galaxies
in clusters provide a means of assessing the amount of matter associated with
clusters.  It is generally observed that the velocity of galaxies approach a
constant value for large distances from the cluster core.  As noted above this
indicates the presence of significantly more mass than is visible in the
galaxies themselves.  A detailed statistical analysis of galaxy
dynamics\citea{dp} yields
\be \OmegaClus = 0.15\pm 0.06. \ee

An independent method of verifying this value of $\OmegaClus$ is due to the
observation of giant luminous arcs by Lynds and Petrosian\bckup.\citea{lp}
These arcs are the image of a bright background object that falls in the line
of sight of a cluster core.  The mass of the cluster serves as a gravitational
lense of this background object producing the arc\bckup.\citea{paczynski}
Although the modeling of the mass distribution in the cluster can be quite
complex, the general prediction of $\OmegaClus \sim 0.2$ is in good agreement
with the above value.

Finally we note that many of these methods can be applied on even larger
scales.  For example, the peculiar velocities of clusters of galaxies can be
studied similar to what has been done for galaxies\bckup.\citea{dekel} The
result of these types of studies is a consistent bound of $\Omega > 0.3$.  To
be conservative we shall adopt
\be \OmegaClus > 0.1. \ee
This limit is shown in figure~\ref{fig:main-fig}.

\subsection{Baryon Content of Clusters}

In addition to providing a measure of $\Omega$, clusters also provide a means
of estimating \OmegaB.  The three main mass components of clusters of galaxies
are star in galaxies, hot intracluster gas, and dark matter.  The first two
are comprised solely of baryons.  Optical observations provide an estimate of
the baryon mass in stars and x-ray studies provide an estimate of the baryons
in hot gas.  These two quantities together provide an estimate to the total
baryon mass in the cluster.  The total mass of the cluster is more difficult
to determine.  It is sensitive to numerous assumptions.  In particular the
cluster is typically assumed to be spherical and in dynamical equilibrium
(virialized).  If either of these assumptions are not valid the derived total
mass could be incorrect.  Frequently structure formation models are employed
to remove some of this sensitivity.  White, Navarro, Evrard, and
Frenk\citea{wnef} employed a ``standard'' cold dark matter~(CDM) model
($\Omega_0=1$) coupled with optical and x-ray studies to deduce a baryon
fraction for the Coma cluster of
\be \OmegaBC = (0.009\pm0.002) + (0.05 \pm 0.01) h^{-3/2}. \ee
Here the first term is due to baryons in stars and the second to baryons in
hot gas.  Note that $\OmegaBC$ is defined by
\be \OmegaBC = \frac{\MB}{M_{\rm tot}}, \ee
where $\MB$ is the mass in baryons of the cluster and $M_{\rm tot}$ is its
total mass.  The region defined by these limits is shown in
figure~\ref{fig:OmegaBC}.

If clusters are a fair sample of the universe then we expect $\OmegaB = \left.
\OmegaBC \right/ \Omega_0$ which is clearly not the case if $\Omega_0 = 1$.
The question of whether galaxies in clusters trace the dark matter is still an
open one.  Babul and Katz\citea{bk} found that baryons in an $\Omega_0=1$ CDM
model are more strongly concentrated than the dark matter.  Thus $\OmegaBC >
\OmegaB$ and there is no inconsistency in the results.  Alternate models with
some admixture of hot dark matter also yield $\OmegaBC > \OmegaB$. At present
there are still a number of difficulties to be worked out in the
interpretation of the x-ray gas in clusters result.  It is clear, however,
that this observation provides important constraints on cluster formation
models.

\subsection{$\Omega_0 = 1$?}

A well known feature of FRW cosmologies is at an epoch $t$ if $\Omega<1$
($\Omega>1$) the universe evolves towards $\Omega=0$ ($\Omega=\infty$) on a
time scale $\sim 1 \left/H(t)\right.$ (see ref.~\citeb{dicke}).  Notice that
$\Omega=1$ is an unstable equilibrium point.  At early times $R(t)$ was
changing rapidly and $H(t)$ was large.  Thus all evolutionary changes occurred
on much shorter time scales.  Since the universe is clearly not more than an
order of magnitude away from $\Omega=1$ today it must have been unity to high
accuracy in all earlier epochs.  In particular we have a good understanding of
the physics of the universe at the beginning of BBN ($t\sim 1\unit{sec}$). If
we trace $\Omega$ to the epoch of BBN we find that $\Omega$ must have been
unity to $\sim 17$ decimal places at that time. The extreme amount of tuning
required to satisfy this is an initial condition within the standard big bang
model.

The theory of inflation succinctly explains this fine tuning, the homogeneity
and isotropy of the universe, and other initial conditions with physics
motivated by particle physics theory.  Most models of inflation require the
universe to be perfectly flat, $\Omega_0= 1$, or at least
$\OMatter+\Omega_\Lambda=1$.  Though this is not a measurement and is an
untested theory it provides a compelling theoretical argument for
$\Omega_0=1$.  Furthermore, recent measurements on the largest
scales\citea{dekel} indicate that $\Omega_0\approx 1$.  Though there is no
firm evidence, we show our bias towards $\Omega_0=1$ by plotting this value as
a dashed line in figure~\ref{fig:main-fig} and~\ref{fig:OmegaBC} and the line
$\OMatter+\Omega_\Lambda =1$ in figures~\ref{fig:Lambda0.4}
and~\ref{fig:Lambda0.8}.

\section {Measurements of $H_0$}

The observational determination of $H_0$ has a long and interesting history
beginning with the original measurement by Hubble\cite{hubble} of $H_0 = 550
\Hunits$.  Since this time the value has been reduced by about a factor of
$\sim 10$ due to a number of systematic errors in the assumptions used by
Hubble.  Currently the measurements fall into two distinct groups: $H_0
\approx 80 \Hunits$ (see ref.~\citeb{Hbig}) and $H_0 \approx 50 \Hunits$
(see ref.~\citeb{sandage}). These two values are represented by dashed lines in
figure~\ref{fig:main-fig}.  At present possible systematic errors do
allow for a consistent resolution of the two values at $H_0\sim 66\Hunits$
which happens to also be near the value recently obtained by Riess, Press, and
Kirshner\bckup.\citea{rpk}  Notice from figure~\ref{fig:Lambda0.4} that
$H_0\sim80\Hunits$ can be made consistent with a flat universe if
$\Omega_\Lambda \sim 0.4$.

The main difficulty in determining $H_0$ is establishing the distance to an
object.  Although it is relatively easy to establish the redshift of an
object, its absolute distance is quite difficult to determine.  The redshifts
for objects observed by Hubble are the same today whereas the distances Hubble
estimated are quite far from present estimates.  The traditional approach for
absolute distance measurements is to identify a standard candle (an object
with a known luminosity) and use its apparent luminosity to determine a
distance.  In practice a given standard candle can only be observed over a
limited distance range.  Thus a ladder of distances to known objects must be
built starting with nearby objects.  Each rung on the distance ladder is
governed by a different standard candle.  Slight errors or disagreements on an
early rung can correspond to large uncertainties and differences for very
distant objects and thus different values for the Hubble constant.

One method of minimizing this problem is to use type~Ia supernovae where the
luminosities are known, at least roughly, from the physics and they can be
observed over a large range of distances. Detailed calibrations of type~Ia
supernovae involves establishing distances by other techniques, including
cepheid variable stars from the Hubble space telescope, to nearby galaxies
where such supernovae have exploded.  This method tends to give a value of
$H_0\sim 50\Hunits$.  Note that this technique can also be used to bound $H_0$
{}from below.  The extreme lower bound comes from the fact that type~Ia
supernovae seem to be caused by a C-O white dwarf star burning its C-O to Fe
via carbon detonation/deflagration.  Assuming that the entire $1.4M_\odot$
Chandrasekhar mass of the white dwarf is pure carbon and is completely
converted to iron provides the maximal energy release and the limit\citea{abw}
\be H_0 \ge 38 \Hunits. \ee
We show this lower limit in figure~\ref{fig:main-fig}.

\section{Two Dark-Matter Problems}

The so called two dark-matter problems are (i)~most of the baryons in the
universe are dark (baryonic dark matter) and (ii)~most of the matter in the
universe is non-baryonic (non-baryonic dark matter).  Both of the problems are
illustrated in figure~\ref{fig:main-fig}.  Note that regardless of the value
of $H_0$, \OmegaVis\ does not intercept \OmegaB.  Similarly on the high
$\Omega$ side, \OmegaB\ only marginally intersects \OmegaClus\ at very low
$H_0$ values.  Thus except for $H_0 < 50\Hunits$ we already know that
\OmegaClus\ requires non-baryonic dark matter.  High values of $H_0$
amplify the need for non-baryonic dark matter.

\subsection{Baryonic Dark Matter}

{}From figure~\ref{fig:main-fig} we see that the regions defined by \OmegaVis\
and \OmegaB\ do not overlap for any value of $H_0$ shown.  Thus even if all of
the visible matter is baryonic, most (at least 70\%) of all baryons must be
dark.  Although the halo of our galaxy could be composed of dark baryonic
objects such as brown dwarfs and Jupiters, known as massive compact halo
objects~(MACHOS), Gates, Gyuk, and Turner\citea{ggt} using the relative
paucity of microlensing events in the direction of the large Magellanic cloud
and the high number of such events towards the galactic bulge have argued that
less than 40\% of the halo can be MACHOS indicating that the halo probably also
includes a non-baryonic component.  This latter point seems to require that at
least some of the non-baryonic dark matter must be cold so it can condense in
halos.  One loop hole to this is black holes with $M\sim 10^3 M_\odot$, too
small to tidally disrupt star clusters and yet big enough to avoid over
production of heavy elements.

\subsection{Non-baryonic Dark Matter}

As noted above, if $H_0 \gtsim 50\Hunits$ the measurement of \OmegaClus\
already requires non-baryonic dark matter.  Moreover, the observational
evidence suggesting $\Omega_0\approx 1$ and the theoretical arguments of
avoiding fine tuning (such as from inflation) requiring a flat universe
further strengthens this requirement.  In the case of $\Omega_0\approx 1$
non-baryonic dark matter is required for all values of $H_0$ since $\OmegaB
< 1$ for any value of $H_0$.  Notice that even for $\Lambda_0 = 0.8$
(figure~\ref{fig:Lambda0.8}) the baryonic and non-baryonic dark matter
arguments are unchanged.

\section{Summary}

We have used the \HOmega\ diagram to illustrate the nature of key problems in
physical cosmology and the role nuclear physics plays in them.  We have seen
how BBN serves to define the two dark matter problems.  We have seen that if
$H_0 > 50\Hunits$ then $\OmegaClus > \OmegaB$ requiring the existence of
non-baryonic dark matter.  If $\Omega_0 = 1$ as current observational and
theoretical work indicates then non-baryonic dark matter is required for all
values of $H_0$.  Finally we have reviewed the age constraints from globular
clusters and nucleocosmochronology to show that $\Lambda_0 =0$ requires $H_0
\le 66\Hunits.$

\section*{Acknowledgements}

We would like to acknowledge our previous collaborators D.~N.~Dearborn,
J.~R.~Gott, J.~E.~Gunn, B.~S.~Meyer, K.~A.~Olive, P.~J.~E.~Peebles,
G.~Steigman, M.~S.~Turner, X.~Shi, J.~W.~Truran, and G.~J.~Wasserburg whose
work we have drawn upon.  We also gratefully remember Beatrice Tinsley who was
the person who brought the GGST collaboration together.  This work has been
supported in part by NSF grant AST 90-22629, DOE grant DE-FG02-91-ER40606, and
NASA grant NAGW-1321 and by a NASA GSRP fellowship.

\beginfigures

\figure{fig:main-fig} The \HOmega\ plane showing allowed and excluded regions.
Shown here are the limits based on the age of the universe, $t_0$, the fraction
of visible matter in the universe, \OmegaVis, the fraction of baryons in the
universe, \OmegaB, the fraction of matter in clusters of galaxies, \OmegaClus,
and type~Ia supernovae, $H_0 \ge 38\Hunits$. Also shown as dotted lines are
the two current values for the Hubble constant, $H_0 = 50\Hunits$ and $H_0=
80\Hunits$, and the theoretically preferred $\Omega_0=1$.  In this figure
$\Lambda=0$.

\figure{fig:Lambda0.4} The \HOmega\ plane showing allowed and excluded regions
as in figure~\ref{fig:main-fig} but with $\Omega_\Lambda=0.4$.

\figure{fig:Lambda0.8} The \HOmega\ plane showing allowed and excluded regions
as in figure~\ref{fig:main-fig} but with $\Omega_\Lambda=0.8$.

\figure{fig:OmegaBC} The \HOmega\ plane showing allowed and excluded regions
as in figure~\ref{fig:main-fig} but with the addition of the limits from
baryons in clusters, \OmegaBC.

\endfigures

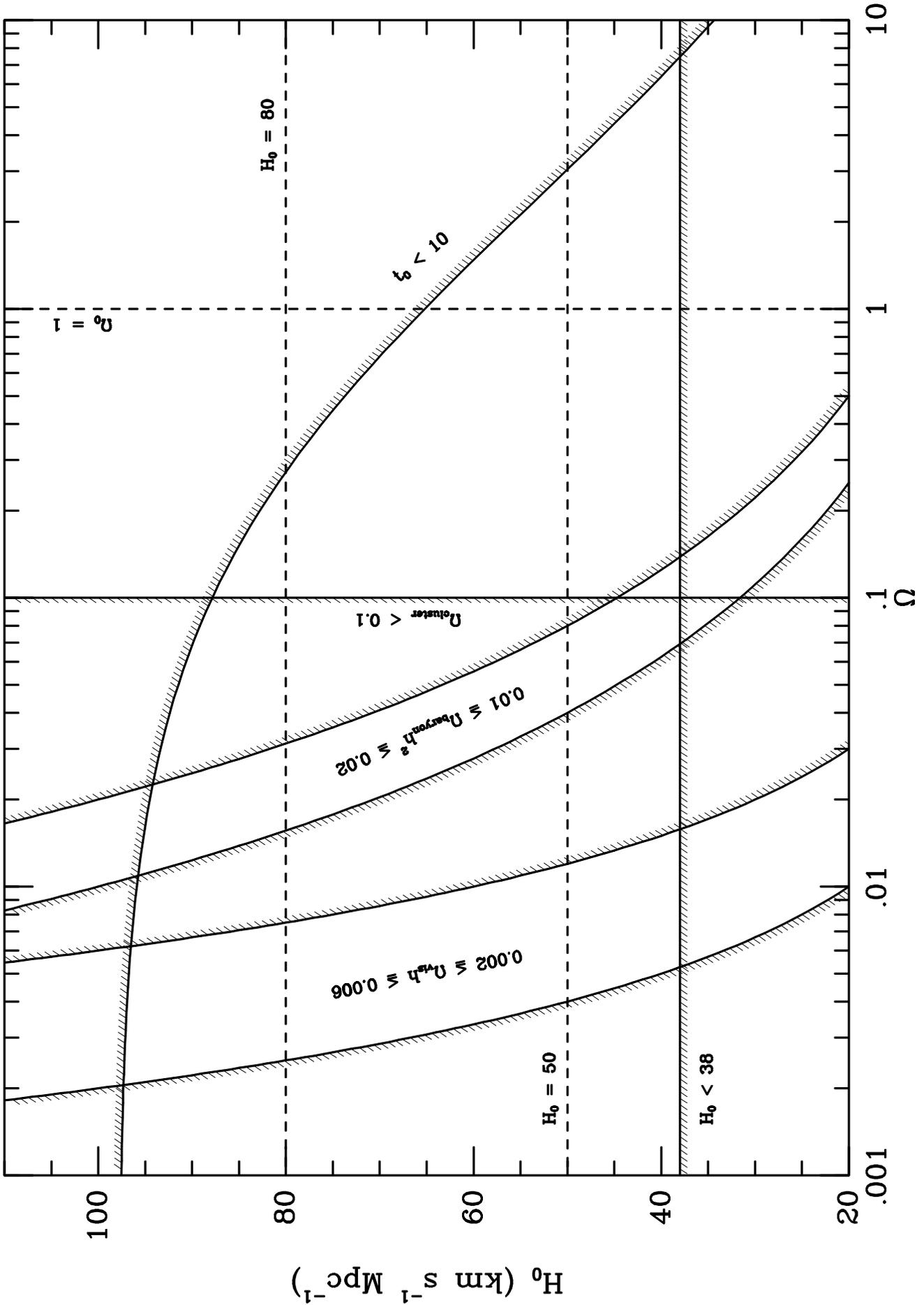
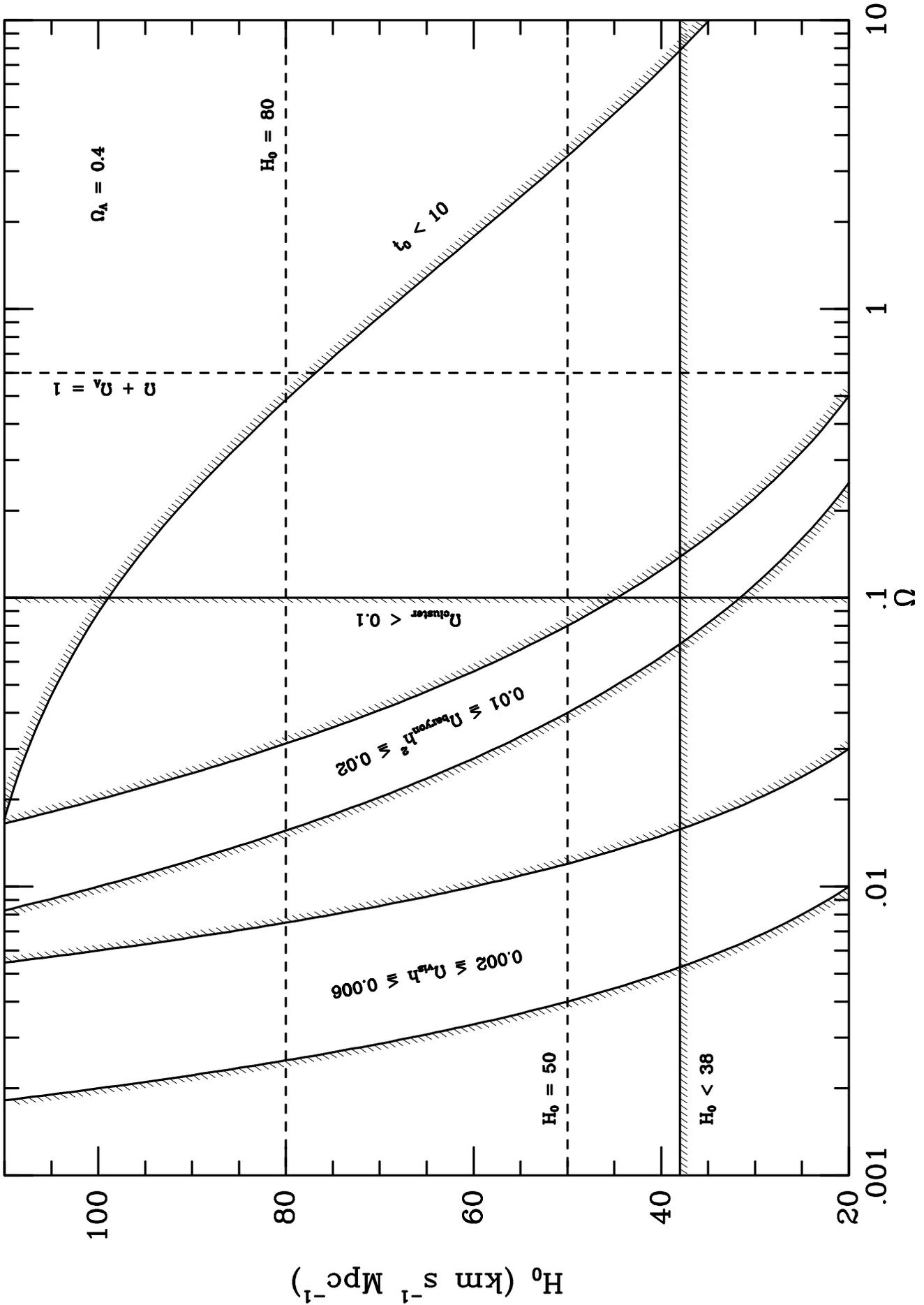
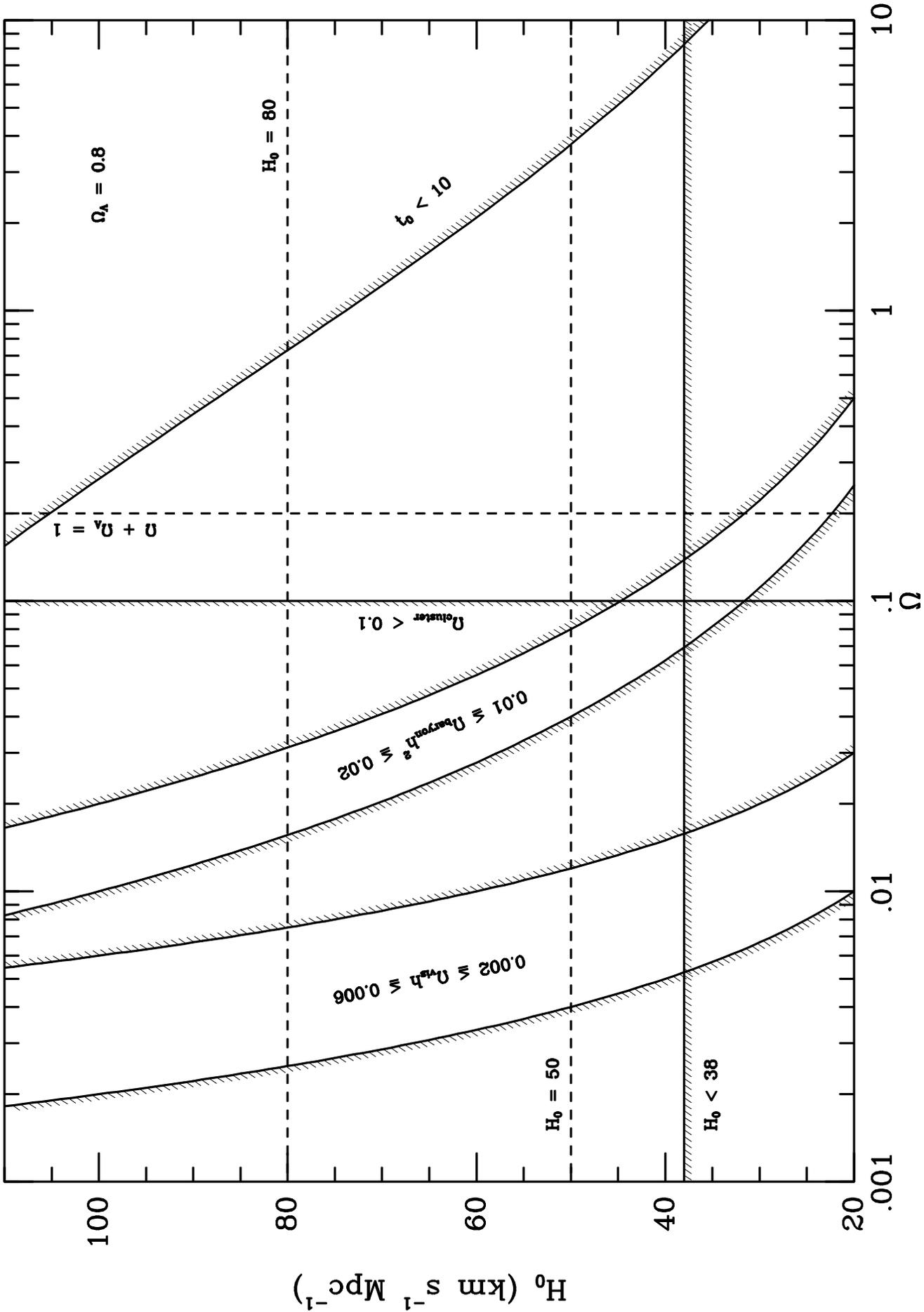
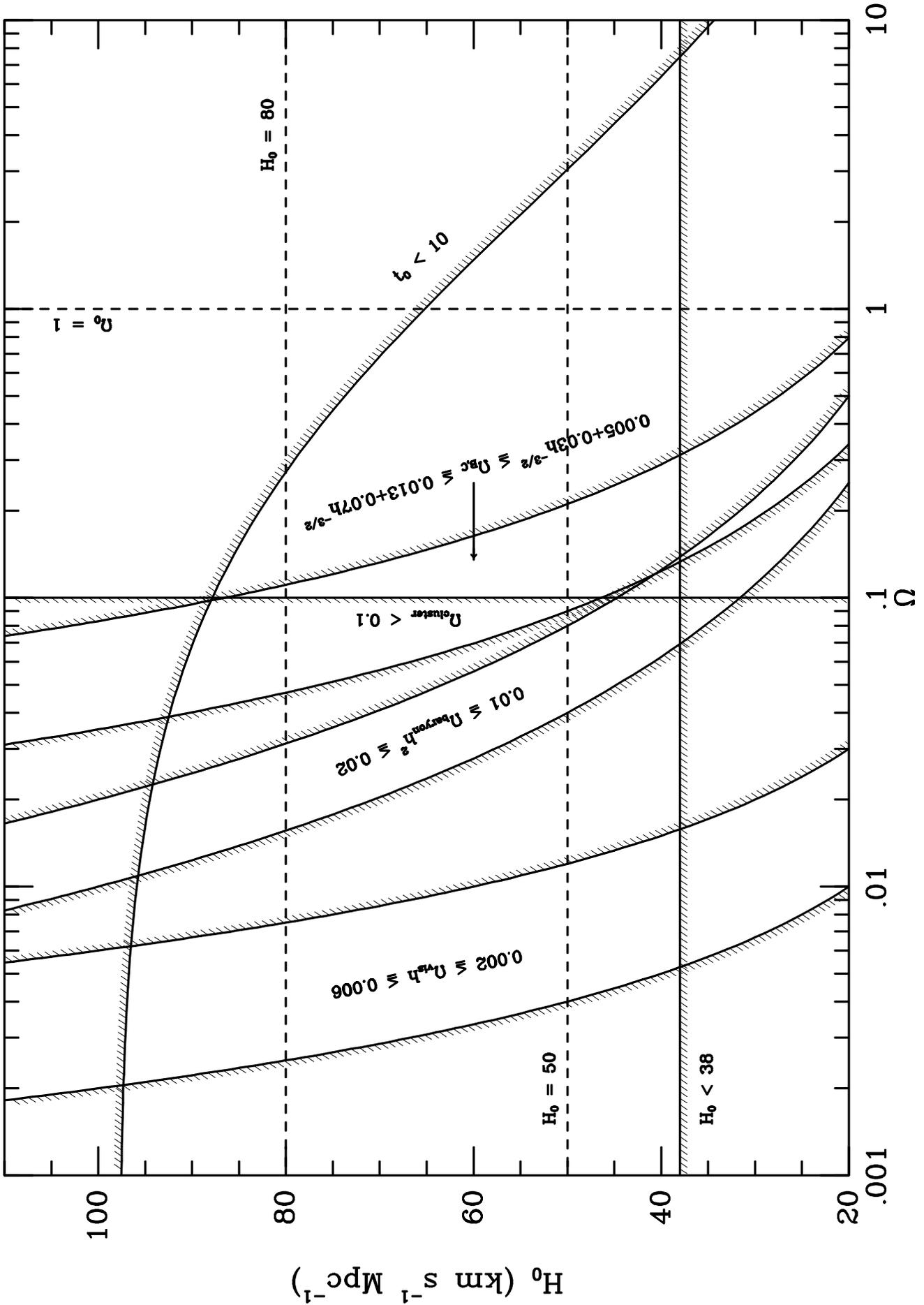
\end{document}